\documentclass[prl,aps,twocolumn,preprintnumbers,showpacs,superscriptaddress]{revtex4}
\usepackage{epsfig,amssymb,amsfonts,verbatim}

\begin{document}
\font\Bbb =msbm10  scaled \magstephalf \def\id{{\hbox{\Bbb I}}}
\newcommand{\ket}[1]{| #1 \rangle}
\newcommand{\bra}[1]{ \langle #1|}
\newcommand{\proj}[1]{\ket{#1}\bra{#1}}
\newcommand{\braket}[2]{\langle #1|#2\rangle}
\newcommand{\half}{\mbox{$\textstyle \frac{1}{2}$}}
\def\opone{\leavevmode\hbox{\small1\kern-3.8pt\normalsize1}}
\newcommand{\tr}[1]{\mbox{Tr} \, #1 }
\def\emph#1{{\it #1}}
\def\textbf#1{{\bf #1}}
\def\textrm#1{{\rm #1}}

\title{Quantum Cryptography Based On Bell Inequalities for Three-Dimensional System}

\author{Dagomir Kaszlikowski}
\author{Kelken Chang}
\affiliation{Department of Physics, Faculty of Science,
National University of Singapore, Lower Kent Ridge, Singapore 119260, Republic
of Singapore}
\author{D. K. L. Oi}
\affiliation{Centre for Quantum Computation, Clarendon Laboratory,
University of Oxford, Parks Road, Oxford OX1 3PU, UK}
\author{L.C. Kwek}
\affiliation{National Institute of Education, Nanyang Technological
University, 1 Nanyang Walk, Singapore 639798}
\author{C.H. Oh}
\affiliation{Department of Physics, Faculty of Science,
National University of Singapore, Lower Kent Ridge, Singapore 119260, Republic of Singapore}

\begin{abstract}

We present a crytographic protocol based upon entangled qutrit
pairs. We analyse the scheme under a symmetric incoherent attack and
plot the region for which the protocol is secure and compare this with
the region of violations of certain Bell inequalities.

\end{abstract}
\maketitle

\section{Introduction}

The need to communicate secretly has always been an important
issue for military strategists during war time. The one-time pad,
first proposed by Vernam, has been shown to be one of the most
secure means of encrypting a message provided the key is truly
random and the key is as long as the message \cite{shannon}.
However, a major problem with the one-time pad is the
establishment of a secure key between the two physically
separated  parties without the services of a courier. Recently,
there has been a major proposal to apply the laws of quantum
mechanics to establish this crucial key. This new proposal,
called Quantum Key Distribution (QKD) protocols, therefore
involves the use of quantum features such as uncertainty
principle or quantum correlations to establish a the necessary
key and hence provides unconditionally secure communication.

The first Quantum Key Distribution was proposed by Bennett and
Brassard (BB84) in 1984 based on the fact that any measurement on
an unknown state of a polarized photon by a third party will
always disturb the state and hence detectable. An extension of the
scheme to three-dimensional quantum states has recently been done
\cite{bruss} and it was shown to be more secure than
two-dimensional case. Another well-known variation of QKD is
based the idea of an entangled pair and detecting the presence of
the eavesdropper using violations of the
Bell-Clauser-Horne-Shimony-Holt (Bell-CHSH) inequality
\cite{ekert91}. This protocol (Ekert protocol) is fundamentally
interesting as it provides an example of how a fundamental
problem in quantum mechanics, namely Bell-CHSH inequality and
violation of local realism, can be applied to a physical problem.
Naturally, one questions if it is possible to extend this latter
protocol involving Bell-CHSH inequality to higher dimensional
system.

The extension of Bell-CHSH inequality to higher dimensions is a
non-trivial and interesting problem.  As higher dimensional
quantum systems require much less entanglement to be
non-separable than two-dimensional systems (qubits), it was
suspected that higher dimensional entangled systems may lead to
stronger violations of local realism. These results have been
shown numerically using linear optimization method by searching
for an underlying local realistic joint probability distribution
that could reproduce the quantum predictions
\cite{KASZLIKOWSKI-PRL-2000} and confirmed analytically
\cite{CHQUTRIT,COLLINS}.

\section{Cryptographic key}

The quantum channel we consider consists of a source
producing two qutrits~\cite{DURT}, which we denote by $A$ and
$B$, in the maximally entangled state
$\ket{\psi} =
\frac{1}{\sqrt{3}} \sum_{k=0}^{2} \ket{k}_A \otimes \ket{k}_B$, where
$\ket{k}_A$ and $\ket{k}_B$ are the $k$-th basis state of the qutrit $A$ and
$B$ respectively (these basis states can represent, for instance, spatial
degrees of freedom of photons). Qutrit $A$ flies towards Alice whereas
qutrit $B$ flies towards Bob. Each observer has at his or
her disposal a symmetric unbiased six-port beamsplitter.

\begin{figure}[h]
\begin{center}
\epsfig{figure=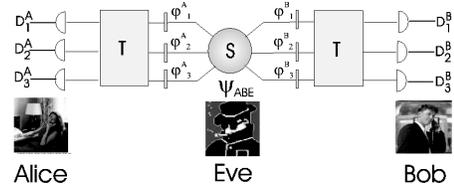,width=0.45\textwidth}
\end{center}
\caption{Qutrit Protocol} \label{fig:protocol}
\end{figure}
An unbiased symmetric six-port beamsplitter performs a unitary
transformation between "mutually unbiased" bases in the Hilbert
space \cite{SCHWINGER60, IVANOVIC81, WOOTERS86}. Such devices were
tested in several quantum optical experiments \cite{MMWZZ95, RECKPHD96}, and also
various aspects of such devices were analyzed theoretically
\cite{RECK94, JEX95}.

This quantum optical device has three input and three output ports. In front
of each input port there is a phase shifter. When all the phase shifters
are set to zero an incoming photon through one of the input ports has an
equal chance to leave the device through any of the output ports.
The elements of the unitary transformation, which describes its
action, are given by
\begin{equation}
U^{k\ell} = \frac{1}{\sqrt{3}} \, \alpha^{k\ell} \, e^{i
\varphi_{\ell}} \,,
\end{equation}
where $\alpha=e^{2 \pi i / 3}$ and the indices $k$, $\ell$
($k,\ell=0,1,2$) denote the input and exit ports respectively; $\varphi_{\ell}$
are the phase shifters. These
phase shifters can be changed by an observer. For convenience, we
will denote the values of the three phase shifts in the form of a
three dimensional vector $\vec{\varphi} =
(\varphi_{1},\varphi_{2},\varphi_{3})$.  In our protocol both
observers perform three distinct unitary transformations on their
qutrits.  The transformations at Alice's side are defined by the
following vectors of phases $\vec{\varphi}^A_1 = (0,0,0)$,
$\vec{\varphi}^A_2 = (0,{\pi\over 3},-{\pi\over 3})$,
$\vec{\varphi}^A_3 = (\pi,0,-\pi)$ whereas the transformations at
Bob's side are defined by $\vec{\varphi}^B_1 = (0,{\pi\over
6},-{\pi\over 6})$, $\vec{\varphi}^B_2 = (0,-{\pi\over
6},{\pi\over 6})$, $\vec{\varphi}^B_3 = (-\pi,0,\pi)$.  The
observers choose their transformations randomly and independently
for each pair of incoming qutrits. After performing the
transformation defined by the vectors of phases
$\vec{\varphi}^A_m, \vec{\varphi}^B_n$ the state $\ket{\psi}$
reads $\ket{{\widetilde{\psi}}}_{mn} =
U_A(\vec{\varphi}^A_m)\otimes U_B(\vec{\varphi}^B_n)\ket{\psi}$.
The observers perform the measurement of the state of the qutrit
in the basis in which $\ket{\psi}$ is defined, that is,
$\ket{0}_x, \ket{1}_x, \ket{2}_x$ ($x=A,B$).  We have adopted an
uncommon but useful complex value assignment to the results of
the measurements, first used in \cite{MMWZZ95}:  namely, for the
result of the measurement of the ket $\ket{k}_x$ we ascribe the
value $\alpha^{k}$.  This value assignment naturally leads to the
following definition of the correlation function
$Q(\vec{\varphi}^A_k,\vec{\varphi}^B_{\ell})$ ($Q_{k\ell}$ for
short) between the values of Alice's and Bob's results of
measurements~\cite{MMWZZ95}
\begin{equation}
Q_{k\ell} = \sum_{a,b = 0}^{2} \, \alpha^{a+b} \, P(a, b;\vec{\varphi}^A_k, \vec{\varphi}^B_{\ell}) \,,
\end{equation}
where $P(a, b;\vec{\varphi}^A_k, \vec{\varphi}^B_{\ell})$ denotes the probability of obtaining the result
$a$ by Alice and the result $b$ by Bob for the respective values of the phase shifts they have used.
It can be shown that the above
correlation function reads
\begin{widetext}
\begin{equation}
Q_{k\ell} = \frac{1}{3} \Big[ e^{i(\varphi_{0}^{A}(k) -
\varphi_{1}^{A}(k)+ \varphi_{0}^{B}(\ell) -
\varphi_{1}^{B}(\ell))} + e^{i(\varphi_{1}^{A}(k) -
\varphi_{2}^{A}(k)+ \varphi_{1}^{B}(\ell) -
\varphi_{2}^{B}(\ell))} + e^{i(\varphi_{2}^{A}(k) -
\varphi_{0}^{A}(k)+ \varphi_{2}^{B}(\ell) -
\varphi_{0}^{B}(\ell))}\Big] \,, \label{corr}
\end{equation}
\end{widetext}
where, for instance, $\varphi_{2}^{A}(k)$ denotes the second
component of the $k$-th vector of phases for Alice.

Note that $Q_{33} = 1$.  This means that the results of the
measurement obtained by Alice and Bob are strictly correlated.
When Alice obtains the results $1,\alpha,\alpha^2$ Bob must
register the results $1, \alpha^2,\alpha$ respectively. Thus,
only the following pairs of the results are possible $\{
(1,1),(\alpha,\alpha^2),(\alpha^2,\alpha) \}$ (denoted
subsequently by $\{(0,0),(1,2),(2,1)\}$) and each pair of
correlations occurs with the same probability equal to ${1\over
3}$.   Let us also define the following quantity
\begin{equation}
S = \mbox{{\rm Im}} (-\alpha^2 Q_{11} + \alpha Q_{12} +\alpha^2
Q_{21} - \alpha^2 Q_{22}) \,. \label{Bell}
\end{equation}
It can be shown~\cite{unpublished}, using the recently discovered
Bell inequality for two qutrits~\cite{CH3}, that according to
local realistic theory $S$ cannot exceed $\sqrt 3$.  However,
when using the quantum mechanical correlation function
(\ref{corr}), $S$ acquires the value ${2\over 3}(2+\sqrt{3})$.
Therefore, to violate the above Bell inequality in this case one
must reduce the correlation function by the factor
${6\sqrt{3}-9\over 2}$ (such reduction is possible by adding the
symmetric noise to the system). It has been proved \cite{genie}
that the above Bell inequality gives necessary and sufficient
conditions for local realism in this case.


After the transmission has taken place, Alice and Bob publicly
announce the vectors of phase shifts that they have chosen for
each particular measurement and divide the measurements into two
separate groups: a first group for which they have used the
vectors $\vec{\varphi}^A_1$, $\vec{\varphi}^A_2$ and
$\vec{\varphi}^B_1$, $\vec{\varphi}^B_2$, and a second group for
which they have used $\vec{\varphi}^A_3, \vec{\varphi}^B_3$.
Subsequently, Alice and Bob announce in public the results of the
measurements they have obtained but only within the first group.
In this way they can compute the value of $S$. If this value is
not equal to ${2\over 3}(2+\sqrt{3})$ it means that the qutrits
have somehow been disturbed. The source of this disturbance can be
either an eavesdropper or noise.  In case of no disturbance the
results from the second group allow them, due to the mentioned
correlations, to generate a ternary cryptographic key. For
instance when Alice gets the sequence of values, say $(1, \alpha,
1, \alpha^2, \alpha^2, 1, \cdots)$ then Bob must get the
following sequence of results, $(1, \alpha^2, 1, \alpha, \alpha,
1, \cdots)$.

\section{Eavesdropping}

Let us consider a symmetric incoherent attack in which the
eavesdropper (Eve) controls the source that produces pairs of
qutrits used by Alice and Bob to generate the cryptographic key.
Naturally, if Eve wants to acquire any information about the key,
she must introduce some disturbance to the state of the qutrits.
Her only chance of being undetected is to hide herself behind what,
to Alice and Bob, may look like an environmental noise in the
channel. We assume that the noise is symmetrical in the sense
that the correlation function in the presence of it reads
\begin{eqnarray}
&&Q_{noise}(\vec{\phi},\vec{\psi}) = V Q(\vec{\phi},\vec{\psi}),
\label{noise}
\end{eqnarray}
where $0 \leq V \leq 1$. This requirement can only be fulfilled
if the reduced state for Alice and Bob (after tracing out Eve's
degrees of freedom) is of the form
\begin{equation}
\varrho_{AB} =
A|\psi\rangle\langle \psi| + B|\chi_1\rangle\langle \chi_1|
+C|\chi_2\rangle\langle \chi_2| + {D\over 9} I\otimes I,
\label{state}
\end{equation}
where the real (not necessarily all positive) numbers $A+B+C+D=1$, and where
the maximally entangled orthogonal states $|\chi_k\rangle$ ($k=1,2$) read
\begin{eqnarray}
|\chi_1\rangle &=& {1\over\sqrt 3}(|00\rangle+\alpha |11\rangle + \alpha^2 |22\rangle)\nonumber\\
|\chi_2\rangle &=& {1\over\sqrt 3}(|00\rangle+\alpha^2 |11\rangle
+ \alpha |22\rangle).
\end{eqnarray}
This choice of states stems from the fact that only the above
states generate correlation functions that are proportional to
$Q(\vec{\phi},\vec{\psi})$. To be more specific, the state
$|\chi_1\rangle$ gives the correlation function $\alpha
Q(\vec{\phi},\vec{\psi})$ whereas the state $|\chi_2\rangle$ gives
the correlation function $\alpha^2 Q(\vec{\phi},\vec{\psi})$.
Thus, if we compute the correlation function on the state
$\varrho_{AB}$, we arrive at the following formula
\begin{eqnarray}
Q_{noise}(\vec{\phi},\vec{\psi})&=& A Q(\vec{\phi},\vec{\psi}) +
\alpha B Q(\vec{\phi},\vec{\psi})
+ \alpha^2 C Q(\vec{\phi},\vec{\psi}) \nonumber\\
&=& (A+\alpha B +\alpha^2 C) Q(\vec{\phi},\vec{\psi}).
\end{eqnarray}
From Eq.(\ref{noise}), we obtain the condition $A+\alpha B
+\alpha^2 C = V$, which is only possible if $B=C$ ($V$ is real).

Eve can prepare the reduced density operator (\ref{state}) by preparing an
entangled state of the form,
\begin{eqnarray}
&&|\psi_{ABE}\rangle = \sqrt{{F\over 3}}(|00\rangle
|E_{00}\rangle +
|11\rangle |E_{11}\rangle + |22\rangle |E_{22}\rangle)\nonumber\\
&&+ \sqrt{{G\over 6}}(|01\rangle |E_{01}\rangle +
|10\rangle |E_{10}\rangle+|20\rangle |E_{20}\rangle\nonumber\\
&& + |02\rangle |E_{20}\rangle + |12\rangle |E_{12}\rangle +
|21\rangle |E_{21}\rangle), \label{ancilla}
\end{eqnarray}
where $\{|kl\rangle\}$ are the computational basis states of the two qutrits,
and $\{|E_{kl}\rangle\}$ are states of ancilla.  Without loss of generality,
we can assume that they are normalized (which implies that $F+G=1$). Note that
the most general state of the joint system of Alice's and Bob's qutrits and
Eve's ancilla reads $\sum_{kl=0}^2 |kl\rangle |E_{kl}\rangle$. However,
Eq.~(\ref{state}) and the requirement that $\varrho_{AB} =
\text{Tr}_{E}(|\psi_{ABE}\rangle\langle\psi_{ABE}|)$
imposes the following
conditions on the states of the ancilla
\begin{eqnarray}
&&F\langle E_{kk}| E_{ll}\rangle =A-B\nonumber\\
&&\langle E_{kl} | E_{mn}\rangle = \delta_{kl}, k\neq l,
\end{eqnarray}
Denoting $\langle E_{kk}|
E_{ll}\rangle$ by $\lambda$ we arrive at the following set of
conditions
\begin{eqnarray}
A+2B+D&=&1\nonumber\\
A-B&=&F\lambda\nonumber\\
D&=&{3\over 2}(1-F).\label{condition}
\end{eqnarray}

Eve's strategy is the following. She prepares the state
(\ref{ancilla}), sends the qutrits to Alice and Bob and keeps her
ancilla. She then waits for public communication between Alice
and Bob. When the settings of Alice's and Bob's apparatus (phase
shifts) are revealed, Eve adopts the following algorithm: (i) If
the chosen settings are not the ones used for the key generation
she ignores the ancilla; (ii) If the settings are the ones for
which the key is generated, i.e., $\vec{\varphi}^A_3,
\vec{\varphi}^B_3$, she identifies the ancilla state.

Let us first find the transformed state in case (ii), i.e., the state
$|\tilde{\psi}_{ABE}\rangle= U_A(\vec{\varphi}^A_3)\otimes
U_B(\vec{\varphi}^B_3)\otimes I|\psi_{ABE}\rangle$. A straightforward
computation yields
\begin{eqnarray}
\ket{\tilde{\psi}_{ABE}}=&&\sum_{a,b=0}^2\ket{ab}\ket{\tilde{E}_{ab}},
\label{eq:transstate}
\end{eqnarray}
where the un-normalized states $|\tilde{E}_{ab}\rangle$ read
\begin{eqnarray}
|\tilde{E}_{ab}\rangle&=&\frac{1}{3}\bigg(\sqrt{\frac{F}{3}}
\sum_{k=0}^2\alpha^{(a+b)k}e^{i(\varphi^A_k(3)+\varphi^B_k(3))}|E_{kk}\rangle\nonumber\\
&+&\sqrt{\frac{G}{6}}\sum_{m\neq n}\alpha^{am+bn}e^{i
(\varphi^A_m(3)+\varphi^B_n(3))} |E_{mn} \rangle\bigg)
\end{eqnarray}
Note that (\ref{eq:transstate}) can also be written more conveniently as
\begin{widetext}
\begin{eqnarray}
|\tilde{\psi}_{ABE}\rangle & = & \left( |00 \rangle
|\tilde{E}_{00}\rangle + |12 \rangle |\tilde{E}_{12}\rangle + |21
\rangle |\tilde{E}_{21}\rangle \right) + \left( |11 \rangle
|\tilde{E}_{11}\rangle + |20 \rangle |\tilde{E}_{20}\rangle + |02
\rangle |\tilde{E}_{02}\rangle \right) \nonumber \\
& & \mbox{\hspace{3cm}} + \left( |22 \rangle
|\tilde{E}_{22}\rangle + |10 \rangle |\tilde{E}_{10}\rangle + |01
\rangle |\tilde{E}_{01}\rangle \right), \label{groupeq}
\end{eqnarray}
\end{widetext}
where we have grouped the terms into three orthogonal subspaces associated
with Alice and Bob generating the correct key $\{(0,0),(1,2),(2,1)\}$, and the
two incorrect keys, $\{(1,1),(2,0),(0,2)\}$ or $\{(2,2),(1,0),(0,1)\}$. Note
also that the ancilla states of one subspace are orthogonal to the ancilla
states of the other subspaces.

\begin{figure}[ht]
\begin{center}
\epsfig{figure=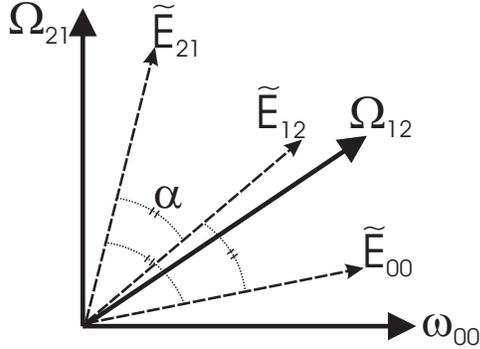,width=0.35\textwidth}
\end{center}
\caption{The optimal three-state discrimination procedure for states in the
  first subspace. The angle between each of the states is
  $\alpha=\arccos{\tilde{\lambda}_1}$.}
\label{fig:discrim}
\end{figure}

The probability that Eve projects into the subspaces spanned by the states $\{
|\tilde{E}_{00} \rangle, |\tilde{E}_{12} \rangle, |\tilde{E}_{21} \rangle \}$,
$\{ |E_{11} \rangle, |E_{20} \rangle, |E_{02} \rangle\}$ and $\{ |E_{22}
\rangle, |E_{10} \rangle, |E_{01} \rangle \}$ are
\begin{eqnarray}
P_{0}&=& 3 \langle \tilde{E}_{00}|\tilde{E}_{00}\rangle=\frac{1+2F\lambda}{3}\nonumber\\
P_{1}&=& 3 \langle \tilde{E}_{11}|\tilde{E}_{11}\rangle=\frac{1-F\lambda}{3}\nonumber\\
P_{2}&=& 3 \langle \tilde{E}_{22}|\tilde{E}_{22}\rangle=\frac{1-F\lambda}{3},
\end{eqnarray}
respectively. We have considered the fact that the states within each bracket
in Eq.(\ref{groupeq}) have the same norms with the same mutual scalar
products.  Moreover, these scalar products are all real.

Eve now has to determine the state of her ancilla, given that
Alice and Bob have projected the whole state into one of three
subspaces associated with the three cases. These subspaces are
orthogonal so that Eve can, in principle, determine without
error, which of these cases Alice and Bob have.

The three ancilla vectors in each subspace corresponding to the result
obtained by Alice and Bob are symmetric and equiprobable.  This makes Eve's
task of discrimination easier as this case has an analytic optimal solution
~\cite{chefles2000} using the so-called ``square-root measurement''. We define
the operator $\Phi=\sum_{ab}\proj{\tilde{E}_{ab}}$, where
$\{\ket{\tilde{E}_{ab}}\}$ are the ancilla states spanning the subspace
associated with Alice and Bob's measurement outcomes.  Since we are
discriminating 3 vectors in a 3-dimensional space, the optimum measurement
directions, $\ket{\omega_{ab}}=\Phi^{-\frac{1}{2}}\ket{\tilde{E}_{ab}}$ are
orthogonal, hence Eve simply performs a projective measurement on her ancilla
(Fig.~\ref{fig:discrim}).

 \begin{figure}[h]
 \begin{center}
 \epsfig{figure=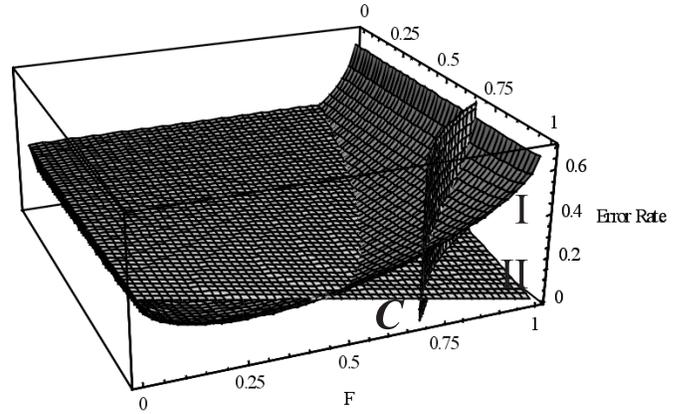,width=0.49\textwidth}
 \end{center}
 \caption{Three Dimensional Plots of the Error Rates}
 \label{fig:error}
 \end{figure}

Thus, Eve's error rate is given by
\begin{equation}
\mathcal{E}_{\text{Eve}}=\sum_{i=0}^3 P_{i} (1-W_i)
\end{equation}
where $W_i, ~ (i=1, 2, 3)$ is the probability of correctly
identifying the three states of the ancilla in the i-th subspace.
These probabilities are given by
\begin{equation}
W_i=\left(\frac{1}{3}\sqrt{1+2\tilde{\lambda}_i}+\frac{2}{3}\sqrt{1-\tilde{\lambda}_i}\right)^2
\end{equation}
where
\begin{eqnarray}
\tilde{\lambda}_1 &= & \frac{1}{2} \frac{3 F + 4 F \lambda -1}{1
+ 2 F \lambda} \\
\tilde{\lambda}_2 &= & \frac{1}{2} \frac{3 F -2 F \lambda -1}{1 - F \lambda}\nonumber\\
&=&\tilde{\lambda}_3
\end{eqnarray}
Due to the symmetry of the noise introduced by Eve, the error
rate between Alice and Bob determined using Eq.(\ref{state}) and
the conditions in Eq.(\ref{condition}) is
\begin{equation}
{\cal E}_{\text{AB}}= \frac{2(1 - F \lambda)}{3}
\end{equation}
We also note that whenever Eve eavesdrops, the correlation function obtained
by Alice and Bob is reduced by $F \lambda$.  Therefore, if this factor is less
than $\frac{6\sqrt{3} - 9}{2}$, the Bell inequality is not violated
\cite{CHQUTRIT} and so Alice and Bob will abort the protocol. This implies
that Eve must keep this factor above this value.

Fig.\ref{fig:error} shows the three dimensional plots of the
error rates of Eve as a function of the parameters $F$ and
$\lambda$ (labeled by surface I) as well as the error rate
between Alice and Bob (labeled by surface II). The region in
which the factor $F \lambda$ is greater than the threshold value
($ V_0=(6 \sqrt{3} - 9)/2$) is demarcated by the ``wall'' labeled
${\cal C}$. In the region bounded by $F \lambda \geq V_0$, the
error rate of Eve is always greater than the error rate between
Alice and Bob.

\begin{figure}
\begin{center}
\epsfig{figure=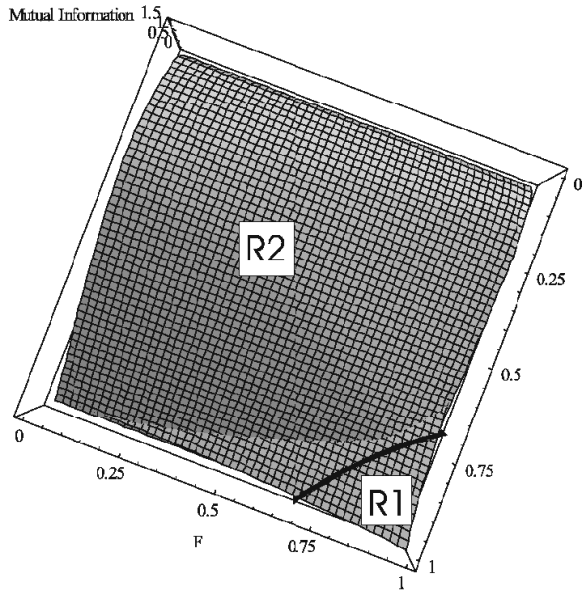,width=0.5\textwidth}
\end{center}
\caption{Plan elevation of 3-dimensional plot of mutual
information between Alice-Bob and Alice-Eve.}
\label{mutual}
\end{figure}

An alternative approach to test the security of the protocol
against such incoherent symmetric attack is to consider the mutual
information between Alice and Eve and compare it with the mutual
information between Alice and Bob. The mutual information between
Alice and Eve is given by the following expression
\begin{widetext}
\begin{eqnarray}
{\cal I}_{\text{AE}} & = & H(A)+H(E)-H(A;E)\nonumber\\
&=&\log 3 - 3 \langle
\tilde{E}_{00} | \tilde{E}_{00} \rangle \log \langle
\tilde{E}_{00} | \tilde{E}_{00} \rangle - 6 \langle \tilde{E}_{11}
| \tilde{E}_{11} \rangle \log \langle
\tilde{E}_{11} | \tilde{E}_{11 } \rangle \nonumber \\
& & -\bigg[- 3 \langle \tilde{E}_{00} | \tilde{E}_{00} \rangle W_1
\log \left(\langle \tilde{E}_{00} | \tilde{E}_{00} \rangle W_1
\right) - 6 \langle \tilde{E}_{00} | \tilde{E}_{00} \rangle (1-
W_1)^2 \log \left(\langle \tilde{E}_{00} | \tilde{E}_{00} \rangle
(1 - W_1)^2 \right)
\nonumber \\
& & - 6 \langle \tilde{E}_{11} | \tilde{E}_{11} \rangle W_2 \log
\left(\langle \tilde{E}_{11} | \tilde{E}_{11} \rangle W_2 \right)
- 12 \langle \tilde{E}_{11} | \tilde{E}_{11} \rangle (1- W_2)^2
\log \left(\langle \tilde{E}_{11} | \tilde{E}_{11} \rangle (1 -
W_2)^2 \right) \bigg],
\end{eqnarray}
where H is the Shannon entropy. The mutual information between Alice and Bob is
\begin{equation}
{\cal I}_{\mbox{\rm \small AE}}  =  2 \log 3 + \frac{1}{3} \left(
1 + F \lambda \right) \left\{ \log \left( 1 + F \lambda \right) -
\log 9 \right \} + \frac{2}{3} \left( 1 - F \lambda \right) \left
\{ \log \left( 1 - F \lambda \right) - \log 9 \right \}.
\end{equation}
\end{widetext}
Fig.~\ref{mutual} shows the plan elevation of the 3-dimensional
plots of the mutual information as a function of the parameters
$F$ and $\lambda$. The line of intersection between ${\cal
I}_{\mbox{\rm \small AE}}$ and ${\cal I}_{\mbox{\rm \small AB}}$
clearly lies behind the wall separating the region in which the
Bell inequality is violated from the region ($R1$) in which local
realistic description is possible ($R2$). In the region $R1$,
${\cal I}_{\mbox{\rm \small AB}}> {\cal I}_{\mbox{\rm \small
AE}}$. From numerical calculation, the maximum value of V for which Eve's
mutual information equals Alice and Bob's is 0.6629. Thus, Alice and Bob have
a buffer region in which to operate securely from this kind of attack by Eve.

To summarize, we have presented a cryptographic protocol using qutrits which
is resistant to a form of symmetric, incoherent attacks. The qutrit Bell inequality
provides a sufficient condition for secure communication. However, this attack
may not be optimal so the Bell inequality may prove to be necessary.

D.K., C.H. Oh and L.C.K. acknowledge financial support provided under
the ASTAR Grant No. 012-104-0040.
D.K.L.O acknowledges the support of CESG (UK) and QAIP grant
IST-1999-11234.

\end{document}